\documentstyle[12pt]{article}
\begin{document}
\setlength{\baselineskip}{5mm}

\noindent{\large\bf

Fourth--Order Ricci Gravity
}\vspace{4mm}

\noindent{

\underline{A. Borowiec}$^a$, M. Francaviglia$^b$ and V.I.
Smirichinski$^c$
}\vspace{1mm}

\noindent{\small
$^a$
 Institute of Theoretical Physics, Wroc{\l}aw University, Poland \\
 $^b$ Departimento di Matematica, Unversit\'a di Torino, Italy \\
 $^c$
 Joint Institute for Nulclear Research,
 Dubna, Moscow  Region, Russia.
}\vspace{4mm}

\def\be{\begin{equation}}
\def\ee{\end{equation}}
\def\bea{\begin{eqnarray}}
\def\eea{\end{eqnarray}}

\def\div{\mbox{{\bf div}}}
\def\cF{{\cal F}}
\def\al{\alpha}
\def\bt{\beta}
\def\ga{\gamma}
\def\Ga{\Gamma}
\def\de{\delta}
\def\ep{\epsilon}
\def\ve{\varepsilon}
\def\ka{\kappa}
\def\la{\lambda}
\def\La{\Lambda}
\def\vr{\varrho}
\def\si{\sigma}
\def\vs{\varsigma}
\def\om{\omega}
\def\Om{\Omega}
\def\th{\theta}
\def\vt{\vartheta}
\def\vp{\varphi}

\def\pr{\prime}
\def\pt{\partial}
\def\na{\nabla}
\def\w{\wedge}
\def\R{\hbox{\bf R}}
\def\J{\hbox{\bf J}}
\def\tr{\hbox{\tt tr}}
\def\det{\hbox{\tt det}}

\begin{abstract}
\noindent The Euler-Lagrange equations of motion for the most
general Ricci type gravitational Lagrangians are derived by means
of a purely metric formalism.
\end{abstract}


Einstein metrics are extremals of the Einstein-Hilbert purely
metric variational problem. It is known that the non-linear
Einstein-Hilbert type Lagrangians $f(R)\sqrt g$, where $f$ is a
smooth function of one real variable and $R$ is the scalar curvature
of a metric $g$ \footnote{One simply writes $\sqrt g$ for
$\sqrt{|\det g|}$.}, lead either to fourth order equations for $g$ which
are not equivalent to Einstein equations unless $f(R)=R-c$
(linear case) or to the appearance of additional matter fields. 

Fourth-derivative theories of gravity enjoy better
renormalizability properties then standard gravity based on the
Hilbert-Einstein Lagrangian density. Besides, higher-order
theories  are of interest in cosmology, since they do contain
every vacuum solution of GR (including Schwarzschild solution).
Linear Ricci type Lagrangians have been studied in this context
by many authors (see e.g. [1, 7-10] and references therein).

In the present  note we are going to use variational
decomposition techniques (see e.g. [4]) in order to obtain  equations
of motion for the most general (nonlinear) Lagrangian densities of
the Ricci type (see formula (6) below) by means of a purely metric formalism.
As an intermediate step one performs the variational decomposition in the
framework of the so-called Palatini formalism, i.e. keeping a metric and
a (symmetric) connection as independent variables [2-6].

In the sequel we shall use lower case letters $r^\al_{\bt\mu\nu}$ and
$r_{\bt\nu}=r^\al_{\bt\al\nu}$ to denote the Riemann and Ricci tensor
of an arbitrary (symmetric) connection $\Ga$:
\bea
r^\al_{\bt\mu\nu} & = & r^\al_{\bt\mu\nu}(\Ga) =
\pt_\mu\Ga^\al_{\bt\nu}-\pt_\nu\Ga^\al_{\bt\mu} +
\Ga^\al_{\si\mu}\Ga^\si_{\bt\nu}-\Ga^\al_{\si\nu}\Ga^\si_{\bt\mu}
\nonumber\\
r_{\mu\nu} & = & r_{\mu\nu}(\Ga)=r^\al_{\mu\al\nu} \eea i.e.
without assuming that $\Ga$ is the Levi-Civita connection of a metric
$g$. We shall assume that space-time is four-dimensional  but our
results will be valid in any dimension $n\geq 3$.

As it is also known, the linear "first order" Lagrangian $r\sqrt
g$, where $r=r(g,\Ga)=g^{\al\bt}r_{\al\bt}(\Ga)$ is a scalar
concomitant of a metric $g$ and  a linear (symmetric) connection
$\Ga$, \footnote{ Now, the scalar
$r(g,\Ga)=g^{\al\bt}r_{\al\bt}(\Ga)$ is no longer the scalar
curvature, since $\Ga$ is no longer the Levi-Civita connection
of $g$.} leads to separate  equations for $g$ and $\Ga$ which
turn out to be equivalent to the Einstein equations for $g$
(so-called Palatini principle, c.f. [1-4]). 
In other words, the first-order metric-affine Lagrangian $r\sqrt
g$ and the purely metric Einstein-Hilbert Lagrangian $R\sqrt g$
are equivalent in the sense that they lead to the same class of
solutions, namely to vacuum Einstein metrics.

Unlike in the purely metric case, an equivalence with General
Relativity also holds for non-linear gravitational Lagrangians \be
L_f(g,\Ga)= \sqrt g\, f(r) \ee (parametrized by the real function
$f$ of one variable) when they are considered within
the first-order Palatini formalism. 
More exactly, it was shown [6] that (with the notable exception
of some special non-generic choice of the function $f$) we always
obtain Einstein equations as gravitational field equations.
This property of Einstein equations was called {\it universality}.

The universality property holds also true for "Ricci squared"
non-linear Lagrangians  \be \hat L_f(g,\Ga)=\sqrt g\, f(s) \ee
where $s=s(g,\Ga)=g^{\al\mu}g^{\bt\nu}r_{(\al\bt)}r_{(\mu\nu)}$
and $r_{(\mu\nu)}=r_{(\mu\nu)}(\Ga)$ is the symmetric part of the
Ricci tensor of $\Ga$ (see [3]). (Thereafter round brackets around
indices denote symmetrization). But in this case, besides the
Einstein equations, one obtains algebraic constraints which can be
interpreted as an almost-product or an almost-complex structure
on the space-time manifold [2]. Moreover, which one of these two
can be realised depends on the sign of a cosmological constant
which is introduced within the calculations (see [3] for details).

Let us consider a $(1,1)$ tensor valued concomitant of a metric
$g$ and a linear torsionless connection $\Ga$ defined by \be
S^\mu_\nu\equiv S^\mu_\nu (g, \Ga)=g^{\mu\la}r_{(\la\nu)} (\Ga)
\ee One can use it to define a family of scalar concomitants of
the Ricci type \be s_k=\tr S^k \ee for $k=1,\ldots, 4$. We can
in fact eliminate the higher order Ricci scalars $s_k$ with $k > 4$
by using the characteristic polynomial equation for the $4\times 4$
matrix  $S$ (see [5]). One immediately recognizes that $r\equiv
s_1=\tr S$ and $s\equiv s_2 =\tr S^2$.

The most general family of non-linear Ricci-type gravitational
Lagrangians in four-dimensions, i.e.: \be L_F(g,\Ga)=\sqrt g\,
F(s_1, s_2, s_3, s_4) \ee is thence parametrized by a smooth
real-valued function $F$ of $4$-variables. This includes the
Lagrangians (2) and (3) as particular cases.

 As already mentioned above we start with the "Palatini variational
principle", i.e. we choose a metric $g$ and a symmetric connection
$\Ga$ 
as independent dynamical variables. Variation of $L_F$ gives then
\be\de L_F  =  \sqrt g\,\de_g F- \frac{1}{2}Fg_{\al\bt}\,\de
g^{\al\bt} + \sqrt g\,\de_\Ga F \ee where  obviously $\de
F=\sum_{k=1}^4 F^\prime_k\,\de s_k$ and $F^\prime_k=\frac{\pt
F}{\pt s_k}$. We see at once that
$$
\de_g s_k= k\,\tr (S^{k-1}\de_g S)=
k\,(S^{k-1})^\si_\al\, r_{(\bt\si)}\,\de g^{\al\bt}
$$
which is clear from $\de s_k=k\,\tr (S^{k-1}\de S)$. Accordingly one
has \be \de_g F= \cF^\si_\al\, r_{(\bt\si)}\de g^{\al\bt} \ee where for
simplicity we have introduced a $(1,1)$ tensor field concomitant
$$ \cF =\sum_{k=1}^4 k\, F^\prime_k\, S^{k-1} $$ In a similar way
one calculates \be \de_\Ga F=\cF^\al_\si\, g^{\si\bt}\,\de
r_{(\al\bt)} \equiv \cF^{\al\bt}\,\de r_{(\al\bt)} \ee where, as
usual,  the metric $g$ has been used to rise and lower indices.
Moreover, as it can be easily checked by a direct calculation,
the (0,2) tensor field
$$\cF^{\al\bt}\equiv \cF^\al_\si\,g^{\si\bt}$$ is already symmetric (see [5]).

Substituting (8) and (9) into equation (7) gives \be \de L_F = \sqrt
g\,[\cF^\si_\al\, r_{(\bt\si)}- \frac{1}{2}F g_{\al\bt}]\,\de
g^{\al\bt}- \sqrt g\, \cF^{\al\bt}\,\de r_{(\al\bt)} \ee Now,
taking into account the so-called Palatini formula
$$
\de r_{(\al\bt)}=
\na_\mu\de\Ga^\mu_{\al\bt}-\na_{(\al}\de\Ga^\si_{\bt )\si}
$$
$\na_\al$ being the covariant derivative with respect to
$\Ga$ and performing the "covariant" Leibniz rule one gets the
variational decomposition formula under the form
\bea \de L_F & = & \sqrt
g\,[\cF^\si_\al\, r_{(\bt\si)}- \frac{1}{2}F g_{\al\bt}]\,\de
g^{\al\bt}- \na_\nu\, [\sqrt g\, (\cF^{\al\bt}\,\de^\nu_\la
- \cF^{\nu\al}\,\de^\bt_\la)]\,\de\Ga^\la_{\al\bt}+ \div \eea
which splits $\de L_F$ into the Euler-Lagrange part and a
boundary term generically denoted  by $\div$ since we shall not exploit
its explicit form here. The boundary term, however, plays an essential
role during the calculation of a canonical energy momentum complex
(see e.g. [4]). This will be therefore investigated in more details elsewhere.

Accordingly, the Euler-Lagrange field equations ensuing from the
Palatini formalism read as follows \bea
\cF^\si_{(\al}\,r_{(\bt)\si)}=\frac{1}{2}F\,g_{\al\bt}\\
\na_\nu\,[\sqrt g\,\cF^{\al\bt}]=0 .\eea Their analysis have been already
performed in [5]. Also for this case the universality of Einstein
equations takes place.

Our goal in the present note is to calculate the Euler-Lagrange
equations in the purely metric framework, i.e. assuming
$\Ga^\la_{\al\bt}$ to be the Levi-Civita connection of $g$. For
this purpose one has to continue our calculations with
$\de\Ga^\la_{\al\bt}$ replaced in (11) by
$$\de_g\Ga^\la_{\al\bt} =\frac{1}{2}g^{\la\si}(\na_\al\de g_{\bt\si}+
\na_\bt\de g_{\al\si}-\na_\si\de g_{\al\bt}).$$ For example,
transvecting the first term above  with the second part of (11)
one calculates:
$$\na_\nu\, (\sqrt g\, ([\cF^{\al\bt}\,\de^\nu_\la
 - \cF^{\nu\al}\,\de^\bt_\la])\,\frac{1}{2}g^{\la\si}\na_\al\de
 g_{\bt\si}=\div - \frac{1}{2}\na_\al\na_\nu\, (\sqrt g\, ([\cF^{\al\bt}\,\de^\nu_\la
 - \cF^{\nu\al}\,\de^\bt_\la])\,\de g_{\bt\si}$$
Similar calculations for the two remaining terms  give rise
finally to the fourth-order Euler-Lagrange equations for the
Lagrangian densities (6) which, after some calculations,  can be
presented in the following form: \be \frac{1}{2}\Box\cF^{\al\bt}
+\frac{1}{2}g^{\al\bt}\na_\si\na_\rho\cF^{\si\rho}-\na_\si\na^{(\al}\cF^{\bt)\si}-
\frac{1}{2}g^{\al\bt}F+\cF^{\si(\al}R^{\bt)}_\si =0\ee where
$\Box$ denotes the d'Alambertian (or Laplace-Beltrami) second
order differential operator and $R_{\mu\nu}$ is the Ricci tensor
of $g$. Since $\cF^{\al\bt}$ depends on the second-order
derivatives of $g$ equation (14) provides a fourth-order
non-linear operator acting on $g$.\medskip\\
\noindent {\bf Example 1.} Take $F=f(R)$. Then $\cF^{\al\bt}=
f^\prime(R)g^{\al\bt}$ and (14) yields the well-known result \be
[g^{\al\bt}\Box-\na^\al\na^\bt]f^\prime (R)-\frac{1}{2}f(R)
g^{\al\bt} +f^\prime(R)R^{\al\bt}
=0\ee\medskip\\
\noindent {\bf Example 2.} Take $F=f(R_{\mu\nu}R^{\mu\nu})$; then
$\cF^{\al\bt}=2 f^\prime(R_{\mu\nu}R^{\mu\nu})\,R^{\al\bt}$. In
this case (14) gives \be \Box(f^\prime R^{\al\bt})+
g^{\al\bt}\na_\si\na_\rho(f^\prime
R^{\si\rho})-2\na_\si\na^{(\al}(f^\prime R^{\bt)\si})-
\frac{1}{2} f(R_{\mu\nu}R^{\mu\nu})g^{\al\bt}+2f^\prime
R^{\si(\al}R^{\bt)}_\si =0, \ee which for the linear case
$f^\prime=1$ reduces to \be \Box R^{\al\bt}+
g^{\al\bt}\na_\si\na_\rho
R^{\si\rho}-2\na_\si\na^{(\al}(R^{\bt)\si})-
\frac{1}{2}g^{\al\bt}R_{\mu\nu}R^{\mu\nu}+2 R
^{\si(\al}R^{\bt)}_\si =0 .\ee This can be recasted (using
Bianchi identities) as follows (compare [9]):\be \Box R^{\al\bt}+
\frac{1}{2} g^{\al\bt}\Box R -\na^\al\na^\bt R -
\frac{1}{2}g^{\al\bt}R_{\mu\nu}R^{\mu\nu}+2 R
^{\bt\si\al\rho}R_{\si\rho} =0 .\ee
\section*{Acknowledgment}
This research was supported by the Bogolyubov-Infeld project and
the Polish Committee for Scientific Research  (KBN): grant 2 P03B
144 19~.


\newcommand{\etal}{{\em et al.}}
\setlength{\parindent}{0mm}
\vspace{5mm}
{\bf References}
\begin{list}{}{\setlength{\topsep}{0mm}\setlength{\itemsep}{0mm}%
\setlength{\parsep}{0mm}}
%
\item[1.] D. Barraco and V.H. Hamity, Gen. Rel. Grav. {\bf 31},
213 (1999).
\item[2.] A. Borowiec, M. Ferraris, M. Francaviglia and I. Volovich,
J. Math. Phys. {\bf 40}, 3446 (1999).

\item[3.] A. Borowiec, M. Ferraris, M. Francaviglia and I. Volovich,
 Class. Quantum Grav. {\bf 15}, 43 (1998).

\item[4.] A. Borowiec and M. Francaviglia,
{\it Alternative Lagrangians for Einstein metrics}, in: Proc.
Current Topics in Mathematical Cosmology,
 Int. Sem. Math. Cosmol. Potsdam 1998,  eds.
M. Rainer and H.-J. Schmidt, WSPC, (Singapore,1998) p.361.
\item[5.]
A. Borowiec,  {\it Nonlinear Lagrangians of the Ricci type}, Rep. Math. Phys.
-- in press, gr-qc/9906043 (2000).

\item[6.] M. Ferraris, M. Francaviglia and I. Volovich,
 Class. Quantum Grav. {\bf 11}, 1505 (1994).

\item[7.]
M. Ferraris, M. Francaviglia and G. Magnano,
 Class. Quantum Grav, {\bf 5}, L95 (1988).

\item[8.]
G. Magnano, M. Ferraris and M. Francaviglia, Gen. Rel. Grav. {\bf
19}, 465 (1987).
\item[9.]
L. Querella,
 {\it Variational Principles and Cosmological Models in Higher-Order Gravity},
Doctoral dissertation, gr-qc/9902044 (1999).

\item[10.] K. S. Stelle, Gen. Rel. Grav. {\bf 9},
353 (1978).

%
\end{list}

\end{document}